\documentclass[journal]{IEEEtran}
\usepackage{xcolor}
\usepackage{subfigure}
\usepackage{graphics} 
\usepackage{epsfig} 
\usepackage{amsfonts,amssymb}
\usepackage{multirow}

\ifCLASSINFOpdf
\else
\fi

\begin{document}
	%
	\title{Overcoming the Channel Estimation Barrier in Massive MIMO Communication Systems}
	%
	%
	%
	\author{Zhenyu~Liu,
        Lin~Zhang, and Zhi~Ding
\thanks{Z. Liu is with the School of Information and Communication Engineering,
Beijing University of Posts and Telecommunications, Beijing 100876, China,
and also with the Department of Electrical and Computer Engineering,
University of California at Davis, Davis, CA 95616 USA (e-mail:
lzyu@bupt.edu.cn).}
\thanks{L. Zhang is with the School of Information and Communication
Engineering, Beijing University of Posts and Telecommunications, Beijing
100876, China (e-mail: zhanglin@bupt.edu.cn).}
\thanks{Z. Ding is with the Department of Electrical and Computer Engineering,
University of California at Davis, Davis, CA 95616 USA (e-mail:
zding@ucdavis.edu).}
}
	\maketitle

	\begin{abstract}
		A new wave of wireless services, including virtual reality, autonomous driving and internet of things, is driving the design of new generations of wireless systems to deliver ultra-high data rates, massive number of connected devices and ultra low latency. Massive multiple-input multiple-output (MIMO) is one of 
		the critical underlying technologies that allow
		future wireless networks to meet these service needs.
		This article discusses the application of deep learning (DL) for massive MIMO channel estimation in  wireless networks by 
		integrating the underlying characteristics of 
		channels in future high speed cellular  deployment. 
		We develop important insights derived from the
		physical radio frequency (RF) channel properties and present a 
		comprehensive overview on the application of DL 
		for accurately estimating channel state 
		information (CSI) with low overhead. We provide examples of successful
		DL application in CSI estimation for massive  MIMO
		wireless systems and highlight several
		promising directions for future research. 
		
	\end{abstract}
	

	%
	\IEEEpeerreviewmaketitle
	\vspace*{-2mm}

	\section{Introduction}

Current and future generations of wireless networks
must cope with the continuous and rapid growth of
of applications and data traffic to
deliver ultra-high data rate over wide
coverage area for 
massive number of connected devices and
to support short-latency and low energy 
applications. One of the most important technical
advances at the radio frequency (RF) physical layer is the emergence of massive multiple-input multiple-output (MIMO) transceivers.
	
	 By exploiting spatial diversity and multiplexing gains, massive MIMO  can help improve spectrum efficiency
	and robustness of wireless communication systems under
	limited bandwidth and channel fading. 
To fully utilize their potential gains,
massive MIMO transmitters require sufficiently 
accurate channel state information (CSI) on the 
forwardlink.  
However, the large number of
antennas and the wide bandwidth in high rate links 
significantly increases CSI dimensionality that 
poses
serious challenges to 
traditional channel estimation and feedback 
techniques.

On the one hand, owing to the large number of antennas in massive MIMO systems, channel estimation suffers from  high signal acquisition cost and large training overhead. On the other hand, different uplink and downlink frequency bands in frequency division duplex (FDD) leads to weaker reciprocity between the
two channels. Consequently, gNB (or gNodeB) 
transmitters in
FDD networks would require user equipment (UE) to provide downlink CSI feedback frequently.  
Because of the larger antenna number and broader downlink bandwidth, traditional approaches for
CSI feedback may consume staggering amount
of uplink channel capacity.
Accordingly, the need for accurate downlink CSI constitutes a serious barrier in future high speed wireless communication systems.
	
Although the problem of downlink CSI estimation and feedback under
	massive MIMO appears to be insurmountable, the physical traits
	of the RF channels can provide insights on how to overcome this
	barrier. Specifically, MIMO channels exhibit a number
	of important physical characteristics including spatial coherence, 
	spectral coherence and temporal coherence.  
	By exploiting these inherent channel correlation characteristics,
	we can substantially improve the efficiency
of CSI estimation and feedback in practical massive MIMO applications. 
	Thus, learning and effectively utilizing such channel characteristics	naturally constitute vital parts of our practical solution to
	downlink CSI acquisition in massive MIMO systems. 
	
To overcome challenges in massive MIMO systems challenges, compressive sensing (CS)
has been studied for channel estimation and channel feedback. CS-based approaches exploit the spatial coherence and channel sparsity that stem from the limited scattering characteristics of signal propagation, and can formulate a compressed representation of CSI matrices. However, there
are also limitations. For example, most CS-based  approaches  impose  the  strong  channel sparsity 
condition 
in some domain which may not hold
exactly. Additionally, the sparsity of CSI matrix 
is not exactly on the channel
sampling grid, which may lead to degraded performance due to the power leakage effect around the recovered discrete CSI samples. Furthermore,
CS-based approaches are often iterative,
which can cause additional computation delay.


Recently, 	
deep learning (DL) has emerged as a powerful tool 
for learning the underlying structures from 
large measurement of data. 
DL has achieved notable
	success in areas including computer vision, natural language processing 
	and decision making. Although still in a nascent stage, DL has recently found several interesting applications in the physical layer of wireless communications \cite{wcindl},  
	including signal 
	detection, channel estimation,  low rate CSI feedback, among others. 
However, how to effectively apply DL techniques to exploit RF
	channel properties remains an open research issue, 
	as
	many existing DL based works do not explicitly utilize the
	physical RF characteristics and provide 
	insufficient physical insights despite apparent successes. 
	
	In this work, we emphasize the importance of physical insights in 
	applying DL for 
	downlink CSI estimation and feedback of massive MIMO wireless links.  
In wireless communications, there exists a wealth of expert knowledge on various channel models for
designing and achieving fast and reliable data links. 
Integrating DL with domain knowledge of RF channels acquired over decades of intense research in wireless networks, such as spatial correlation, temporal correlation, and spectral coherence, can provide important insights to clear the CSI barrier 
for massive MIMO systems.

	
In typical cellular systems,
uplink and downlink channels form a bi-directional
RF link with characteristics depending on
physical attributes such as bandwidth, wavelength, 
multipaths and scatters. These physical characteristics 
lead to CSI correlations often captured in 
three domains:
	
	\begin{itemize}
		\item \textbf{Spatial and spectral correlation:} 
		Since RF channels of adjacent sub-carriers or
		adjacent antenna elements 
		exhibit similar propagation characteristics, MIMO CSI
		should be spatially and spectrally correlated. 
		For example, 
		the downlink channel responses are dependent of the propagation 
		gains and the angles associated with
	primary reflectors and scatters. Such intrinsic channel coherence 
		can simplify CSI estimation in massive MIMO
to require fewer observations and less feedback. 
		Through 
		supervised learning, DL algorithms can acquire
		such spatial and spectral coherences for 
		effective estimation of downlink CSI 
		from partial pilots or compressed CSI feedbacks from user equipment (UE). 
		
		\item \textbf{Temporal correlation:} 
	It is well known that even for mobile environment
		under severe Doppler effect, 
RF channel responses are temporally correlated in
		typical massive MIMO configurations. 
		As a result, temporal correlation of massive MIMO channels
	can be exploited to reduce the amount of pilots and  UE feedback
		in downlink CSI estimation. 
		
		\item \textbf{Bi-directional correlation:} 
		Traditionally in TDD systems, an
		uplink MIMO CSI can approximate its corresponding downlink CSI 
based on channel reciprocity. For FDD systems, however,
	such channel reciprocity weakens because uplink
		and downlink frequency bands differ. 
		Still, FDD uplink and downlink channels experience the same RF
		environment. In fact, existing works have demonstrated 
		bi-directional correlation between the two 
		channels in terms of directionality
		\cite{related1}, shadowing effects \cite{related2}, multipath delays \cite{related3}, channel covariance \cite{related6}, etc. 
		Hence, bi-directional CSI correlation allows us to exploit
		uplink CSI at gNB
		to improve downlink CSI estimate by
reducing UE feedback of downlink CSI
estimate. 
		
	\end{itemize}

Based on these special characteristics of physical wireless channels,
appropriately designed DL architectures and algorithms 
can potentially help reduce the amount of
UE feedback for CSI estimation in massive MIMO links. 
Clearly, how to configure, adapt, and improve such tools for accurate CSI estimation by
massive MIMO gNB represents an important technical
barrier, as well as an exciting and promising 
research issue. 
In this article, we present an overview	on the integration of DL in massive MIMO systems, 
highlight some promising results, and 
outline some future research directions.

\vspace*{-2mm}
	\section{Current Works on DL for CSI Estimation}
	
	In this section, we introduce the basics of several relevant
	DL neural networks in wireless communications.
	
	
	\begin{figure*}[bt]
		\centering
		\includegraphics[scale=0.35]{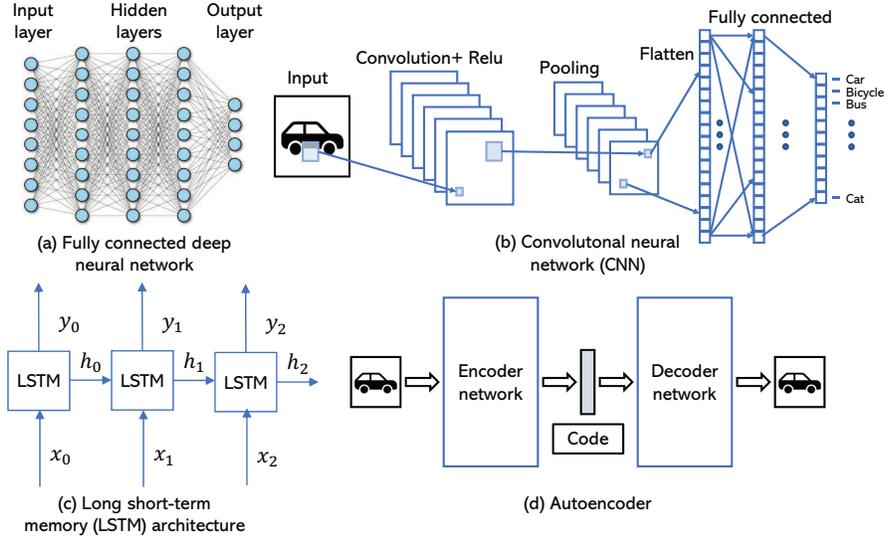}
		\vspace*{-1mm}
		\caption{Commonly utilized deep learning architectures.} \vspace*{-3mm}
		\label{figuren4}
	\end{figure*}
	\subsection{ Fully Connected Deep Neural Network}
Fully connected deep neural networks can
	extract appropriate features for classification and regression, as shown in Fig. \ref{figuren4}(a).
Starting by sending measurement data to
the input layer, each successive layer attempts 
feature extraction from the input data, 
gradually accentuating features that
affect decision making while 
suppressing irrelevant features. 
Through optimization of network parameters, 
DL can be trained to capture underlying data
structures and models despite outliers and noises.  
	
\subsection{Convolutional Neural Networks (CNNs)}
CNNs have been widely applied in problems such as 
image analysis. As shown in Fig. \ref{figuren4}(b), CNNs specialize in processing 
data with grid-like structures and 
include special layers for functions 
such as convolution and pooling.  By stacking convolutional and pooling layers 
alternately, CNN can progressively learn 
complex models.
	
Considering the correlation in spatial, temporal, or 
spectral domains, CSI matrices of massive MIMO systems 
can often be viewed as two-dimensional images.  
CNNs have strong potentials for success 
in massive MIMO channel estimation.

\subsection{ Recurrent Neural Network (RNN) and Long Short-Term Memory (LSTM)}
	RNN is a class of neural networks that exploit sequential information
	by using earlier outputs as part of inputs
	in later time. 
	Unlike traditional neural networks, RNNs 
use internal 
	state (memory) to store
 information that has been calculated.
	An RNN consisting of LSTM units is often 
known as an LSTM network.  
LSTM networks can handle
exploding and vanishing 
gradients in traditional RNNs, 
and work well in classification and 
decision-making according to time series data.
	
Exploiting the temporal correlation of CSI, RNN and LSTM 
can further improve the accuracy of CSI estimates.
	
	\vspace*{-2mm}
	\subsection{Autoencoder}
	
An autoencoder is a neural network trained  
to efficiently
	regenerate its input.
Shown with the structure of Fig. \ref{figuren4}(d), modern autoencoders have generalized the idea of an encoder and a decoder beyond deterministic functions to stochastic mappings. 
An autoencoder can acquire compressed but robust representations of its input and can be
highly effective and efficient in dimension 
reduction or feature learning. 
	
From a DL perspective, CSI feedback within a
wireless communication system can be viewed
as a particular type of autoencoder, which 
aims to recover the
downlink CSI at the gNB side
based on its received CSI that was
compressed by the UE and sent in
the UE feedback. 
	
	
	
	
	\section{Exploiting Wireless CSI Correlations}
Learning and exploiting CSI correlations in massive MIMO can substantially benefit 
CSI compression and recovery for massive MIMO.
	In our recent work \cite{dualnet}, we have presented
two DL-based CSI feedback solutions for massive MIMO wireless communications. We achieved high efficiency by exploiting the underlying CSI correlation 
between uplink and downlink.
Stimulated by this and other preliminary successes, 
we investigate the hidden CSI data structure in terms of spatial and spectral correlation, temporal correlation, and bi-directional correlation of CSI 
for dimension
reduction in massive MIMO wireless applications.
	
	\vspace*{-2mm}
	\subsection{Spatial and Spectral Correlation}
	
	Spatial and spectral correlation of CSI have been characteristics commonly exploited 
	in CS-based feedback and CSI estimation.
Physical RF propagation elements
such as multipaths and scatters provide the 
	foundation of spatial and spectral correlations.  
	For a given gNB, channels are
governed by cell specific attributes such as the buildings, ground and vehicles. Multipath propagation channels and scatters 
between gNB and UE describe how RF signals navigate the path among potential 
obstacles between transmitter and receiver. 
	
Spatial correlation has been experimentally verified \cite{spatial_correlation1}. Smaller antenna separation leads to higher
spatial correlation in massive MIMO.
Since the physical size of antenna arrays at gNB is 
small compared to path distance, paths between different gNB-UE antenna pairs share some common properties \cite{sparse}. On the other hand,
spectral coherence measures channel similarity 
across frequency.
For sub-carriers within channel 
coherence bandwidth, their channels exhibit strong correlation. 
	
Unlike the traditional approaches that often require
reasonably accurate correlation models,  DL can learn and
exploit the underlying channel correlation structure. In
CSI estimation, 
DL algorithms can uncover the inherent but
complex CSI relationship in massive MIMO systems
and can effectively estimate the CSI of many
antennas and sub-carriers 
based on small number of channel sounding pilots. 
For feedback efficiency, DL can also compress high-dimensional CSI to improve the feedback efficiency. 
	
	\subsection{Temporal Correlation}
	
RF channels of mobile UEs are governed by physical scatters, multipaths,
bandwidth, and Doppler effect. For most practical
cases, CSI varies slowly in many massive MIMO systems.
	For mobile users, coherence time 
measures temporal channel 
variations and describes the
Doppler effect caused by UE mobility.
	
Since gNB and UE can store their 
previous CSI estimates, 
UE can encode and feed back the CSI variations,
instead of the full CSI. The
gNB can combine the new feedback with its
previously estimated CSI within coherence time 
for subsequent CSI reconstruction. 
The temporal correlation feature is
a good match with the
capabilities of RNN and LSTM that are
effective in sequential data processing.
	
\subsection{Bi-directional Correlation in TDD and FDD}
	
Historically, the uplink-downlink channel reciprocity has predominantly
been utilized by TDD gNB to infer downlink CSI from 
its own uplink CSI estimate. 
For FDD systems, however, RF components that positively
superimpose in one frequency band may cancel each other in another. 
Hence, FDD uplink-downlink channels do not exhibit 
direct
reciprocity. Nevertheless, because of the shared propagation
environment, correlation still exists between the two.
For example, the angles of arrival of signals
in the uplink transmission are almost the same as the angles of 
departure of signals in the downlink transmission.
When the band gap between uplink and downlink 
channels is moderate, 
both links should share similar propagation characteristics including scattering effects. Such correlation
can be exploited by gNB for CSI estimation in massive MIMO.

Existing works have demonstrated  certain level of
correlation between bi-directional channels for FDD systems.  
The directional 
properties of uplink and downlink FDD channels have been shown 
as correlated \cite{related1}. 
In \cite{related2}, the correlation of shadowing effects between
FDD channels has also been established. In fact,
downlink channel covariance estimation can also
benefit from the observed uplink covariance \cite{related6}. 
Similarly, 
although channel responses could vary for different
downlink and uplink frequency bands, their multipath delays  
remain physically the
same \cite{related3}.  
Thus, to utilize the bi-directional channel correlation, 
channel response matrix from frequency domain should be transformed to
delay domain using inverse Fourier transform.
Compared with the CSI in frequency domain, the reciprocity in delay domain is 
evident and strong owing to 
the shared multipath delays. 


\begin{figure}[!t]
      \centering
      \includegraphics[scale=0.55]{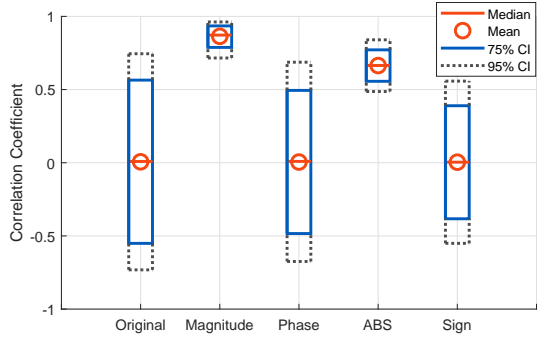}
      \vspace{-1.5ex}
      \caption{Distribution of correlation coefficient between
uplink and downlink CSI at various levels
of Confidence Interval (CI).}\vspace*{-5mm}
      \label{figure1}
\end{figure}

To demonstrate bi-directional correlation, 
we illustrate the correlation coefficient
between uplink and downlink FDD channels obtained through 
numerical tests in Fig. \ref{figure1}. 
We used a pair of transmitter and receiver
in COST 2100 channel model \cite{c2100} to generate
$5.1$ GHz uplink and $5.3$ GHz downlink channel responses. 
As shown in Fig. \ref{figure1}, the correlation coefficients 
between uplink CSI and downlink CSI using the ``original''
(real/imaginary) format are quite erratic. 
Since the CSI is complex-valued, 
their real part and imaginary part correlations are evaluated. 
This test results appear to show weak correlation 
between
corresponding downlink-uplink channel responses.

A closer examination of the physics
reveals that in FDD, CSIs
of two carriers of different 
frequencies may have uncorrelated phases. 
However,
based on FDD multipath channel models,
the CSI magnitudes in delay domain should exhibit much stronger
correlation. To confirm,
we transform the CSI elements
into polar coordinate to separately consider their magnitude
and phase correlations. 
Fig. \ref{figure1} shows that the corresponding
magnitude correlation between uplink and downlink
CSIs exhibit strong correlation whereas their corresponding phases show
very weak correlation. 
In fact, even by removing the signs from the
CSI's real and imaginary parts, 
Fig. \ref{figure1} shows that the absolute values (ABS) 
of uplink and downlink CSI coefficients
are also strongly correlated. However, their 
signs show little correlation. 

These results demonstrate some shared
characteristics 
between uplink and downlink channels in the delay domain. 
This observation provides the basic principle for utilizing
magnitude correlation between uplink and downlink channels in
delay domain for estimating CSI of massive MIMO systems. 

\begin{figure} \centering 

\includegraphics[scale=0.50]{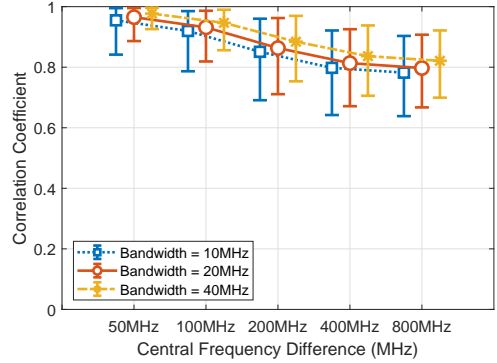}

\caption{Influence of band gap and bandwidth on channel magnitude correlation within 95\% CI.} 
\label{figuren2} \vspace*{-5mm}
\end{figure}

To better understand the correlation between uplink and downlink CSI, 
we evaluated the effect of band gap and bandwidth on bi-directional CSI correlation.
To show the influence of band gap, we keep the central downlink frequency at $5.3$ GHz, while increasing the 
central frequency difference between downlink 
and uplink from $50$ MHz to $800$ MHz. As expected, the results in Fig.~\ref{figuren2} show
that the bi-directional CSI correlation 
weakens with growing band gap. 
To demonstrate the effect of bandwidth on bi-directional
CSI correlation, 
we increase the channel bandwidth from $10$ MHz 
to $20$ MHz and $40$ MHz, respectively, for 
different band gaps.  As 
Fig. \ref{figuren2} shows, 
the bi-directional CSI correlation strengthens for larger channel bandwidth. 
Consequently, 
decreasing the band gap between uplink and downlink 
and increasing the FDD channel bandwidth are two 
ways to strengthen bi-directional CSI
correlations.

\section{DL-Based CSI Estimation in FDD Networks}


In this section, we discuss the applications of 
underlying 
channel characteristics in designing and improving DL-based 
channel estimation and feedback systems.


\subsection{Channel Estimation}
Massive MIMO utilizes large transmit antenna arrays to 
achieve high data rates and multi-user coverage.
It has been viewed as an important technique 
for future wireless system. Channel estimation for massive MIMO system is highly
challenging, especially when the antenna 
array is large and the number of
transceiver RF chains is small.

To overcome the channel estimation challenge in massive MIMO systems, a 
learned denoising-based AMP (LDAMP) estimation method has been proposed for 
the beamspace mmWave massive MIMO system with the help spatial and spectral correlation in \cite{est1}. 
LDAMP incorporates
a denoising CNN into the AMP algorithm for CSI estimation by regarding the CSI matrix
as a 2D natural image. 
LDAMP leverages compressive signal recovery model and utilizes DL network in
iterative signal recovery. This network 
implicitly makes
use of spatial and spectral correlation by
learning the CSI 
structure from a large number of training data.  
LDAMP network was shown to outperform compressive sensing algorithms even when the receiver only has a small number of RF chains.

Another key challenge of CSI estimation in massive MIMO systems is pilot contamination, which stems from interference of pilot symbols utilized by the users in neighboring cells. In \cite{est8},  a DL-based channel estimation method was proposed for multi-cell interference-limited
massive MIMO systems. The proposed estimator employs a deep image prior CNN, designed for image denoising and inpainting, to denoise the received signal first. A conventional least-squares (LS) estimation is then utilized for CSI estimation. Simulation results
show that this deep CSI estimator outperforms
traditional
LS and minimum mean square error (MMSE) estimators 
which are unaware of pilot contamination. 
By testing against uncorrelated Rayleigh
fading channel for each subcarrier, 
the deep CSI estimator no longer performs well,
which shows that its performance gain is from 
exploiting CSI correlations.
 
In massive MIMO, the hardware cost
and power consumption from radio frequency (RF)
chains are also challenging. To  achieve the tradeoff
between the cost and the performance,  mixed analog-to-digital converters
(ADCs) massive MIMO have also been investigated
where a portion of antennas are equipped with high-resolution
ADCs while others employ low-resolution ADCs. In \cite{est6}, with help of spatial and spectral correlation, DL networks are developed to map the CSI from the channels using high-resolution ADCs to those
using low-resolution ADCs. 
Numerical results show that DL-based CSI estimation in mixed-resolution ADCs massive MIMO outperforms linear minimum mean-squared error (LMMSE) methods and expectation maximization (EM) algorithms
based on generalized approximate message passing, especially when using mixed one-bit ADCs.

CSI estimation is more difficult in high-speed 
applications due to the fast time-varying
and non-stationary channel characteristics. In \cite{est7}, a DL network is proposed to tackle the
weaknesses of traditional channel estimation methods in
high-speed mobile scenarios. Specifically, CNN is used to  extract channel response
feature vectors, followed by a RNN for CSI
estimation.  Its simulation results
show that the proposed CSI estimation method
can achieve significant performance
improvement over least squares and LMMSE methods
in high-speed mobile scenarios.




\subsection{Channel Feedback}
In FDD systems, there often exists
a weaker reciprocity between uplink and downlink
channels in different frequency bands. Consequently, UE
feedbacks are often required to report downlink CSI. For massive MIMO,
such feedback data can be substantial because of the large antenna number
and wide bandwidth. Moreover,
in rapidly changing environment, UEs have to feed back CSI frequently.  
Thus, conventional methods based on UE feedback face several challenges, 
including inaccurate channel model and high bandwidth consumption
for UE feedback. 

To preserve feedback bandwidth and improve CSI recovery accuracy, channel properties including spatial, spectral, temporal 
correlation and bi-directional channel correlation may be integrated together with DL-based UE 
feedback. These works have shown substantial performance improvement for downlink CSI estimation in FDD systems with limited feedback resources on uplink.




\begin{figure*}[thpb]
      \centering
      \includegraphics[scale=0.39]{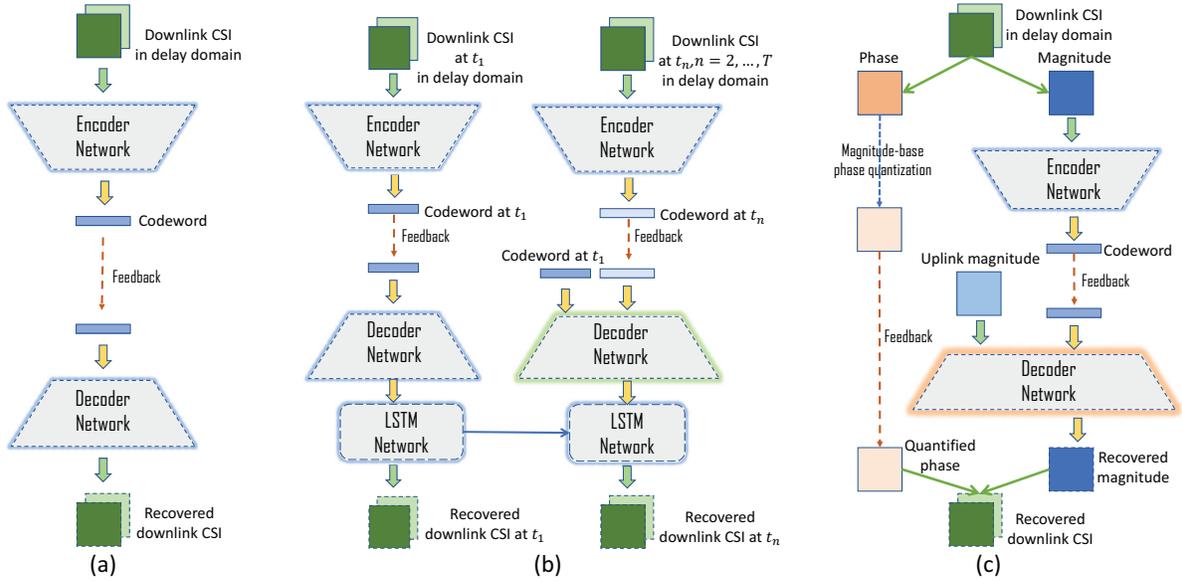}\vspace*{-1mm}
      \caption{CSI feedback architectures. (a) Architecture of CsiNet in \cite{dls}. (b) Architecture of CsiNet-LSTM in \cite{fdtime1}. (c) Architecture of DualNet-MAG in \cite{dualnet}.}
      \label{figure4}
      \vspace*{-3mm}
  \end{figure*}

The authors of \cite{dls} proposed a DL-based CsiNet, as 
illustrated in Fig. \ref{figure4}(a),  to reduce UE feedback overhead 
in massive MIMO systems.  CsiNet mainly utilizes an autoencoder architecture 
that uses an encoder for CSI compression and a decoder for CSI reconstruction. 
CNN is used in both encoder and decoder to exploit 
the spatial and spectral correlation of CSI matrices, 
in a way similar to image processing. The CSI matrices 
are separated to real and imaginary parts,
which correspond to the two sets of input in this neural network. 
CsiNet shows substantial performance gain and  efficiency over some compressive sensing 
methods.

The authors of \cite{fdtime1} adopted LSTM networks to 
exploit temporal channel correlation and lower feedback overhead by designing
CsiNet-LSTM for time-varying massive MIMO channels.
As shown in Fig. \ref{figure4}(b), CSI feedback in CsiNet-LSTM  uses
a sequence within the coherence time. Compared with CsiNet, 
the LSTM network in CsiNet-LSTM is adopted at processing sequence data 
to extract the temporal relationship therein.
For CsiNet-LSTM, only the first MIMO CSI matrix of the time sequence 
is compressed under a moderate compression ratio (CR) and reconstructed by CsiNet.
The ensuing CSI matrices are encoded at a high compression ratio 
by exploiting the temporal correlation. In this way, the authors achieved
a better compression ratio and reduced average feedback payload.  


In \cite{dualnet}, we proposed a DL-based CSI feedback framework
to exploit bi-directional channel correlation characteristics. 
Unlike CsiNet and CsiNet-LSTM, the DualNet in \cite{dualnet} 
exploits the available uplink CSI at gNB 
to help estimate the downlink CSI from low rate UE 
feedback in massive MIMO systems.  We designed two
DL architectures, DualNet-MAG 
and DualNet-ABS, to significantly reduce the UE feedback payload.  Both
DualNet-MAG  and DualNet-ABS can utilize the bi-directional channel correlation
of the magnitude and the absolute value of the CSI coefficients, respectively. 
As shown in Fig. \ref{figure4}(c), the decoder in DualNet-MAG 
reconstructs the downlink CSI magnitudes
based on the uplink CSI magnitudes and its received UE feedback
codewords. 
We further developed a magnitude-dependent phase quantization 
method to reduce the UE phase feedback overhead.  Our work in \cite{dualnet}
shows significant performance gain by DualNet over
other DL architectures relying only on UE feedback.

\section{Open Issues}
We have demonstrated the benefits of integrating 
DL to exploit inherent channel correlations for 
downlink CSI estimation in massive MIMO wireless communications.
We now discuss several future research directions 
to further improve DL-based CSI estimation for future wireless networks. 

\subsection{DL under Different System Settings}
In wireless communications, various system parameters can
can potentially affect the efficacy of DL-based CSI methods. 
As DL techniques are poised to play a bigger role in future wireless systems, 
one important work is to assess how effective DL-based architectures can be
under different network settings. Investigating
the impact of system parameters on DL-based solutions can
provide better design insights for future development.

We can further investigate the influence of bi-directional channel band gap 
and RF channel bandwidth on the performance of DL-based CSI estimation in FDD systems. 
In Section III.C,
we have shown that lower bandgap and larger bandwidth lead to stronger 
bi-directional correlation. One naturally wonders whether they can
improve the performance of CSI estimation.
We can test the performance of CsiNet, DualNet-MAG,
and DualNet-ABS in different bandwidths and bandgaps. 
The central downlink frequencies are set to $5.3$ GHz and $930$ MHz for
indoor and outdoor scenarios, respectively.  
For indoor scenario, the bandgap of 180 MHz and bandwidth of 20 MHz are 
selected as the baselines for comparison. For outdoor scenario, 
the bandgap of 75 MHz and bandwidth of $5$ MHz are set as the baseline. 
We compare the downlink CSI estimates under feedback compression
ratios of 1/8, 1/12, and 1/16. Smaller ratio implies higher compression
in CSI feedback and is more efficient.

\begin{figure} \centering 
\subfigure[Indoor] {\label{fign6:a} 
\includegraphics[width=0.46\columnwidth]{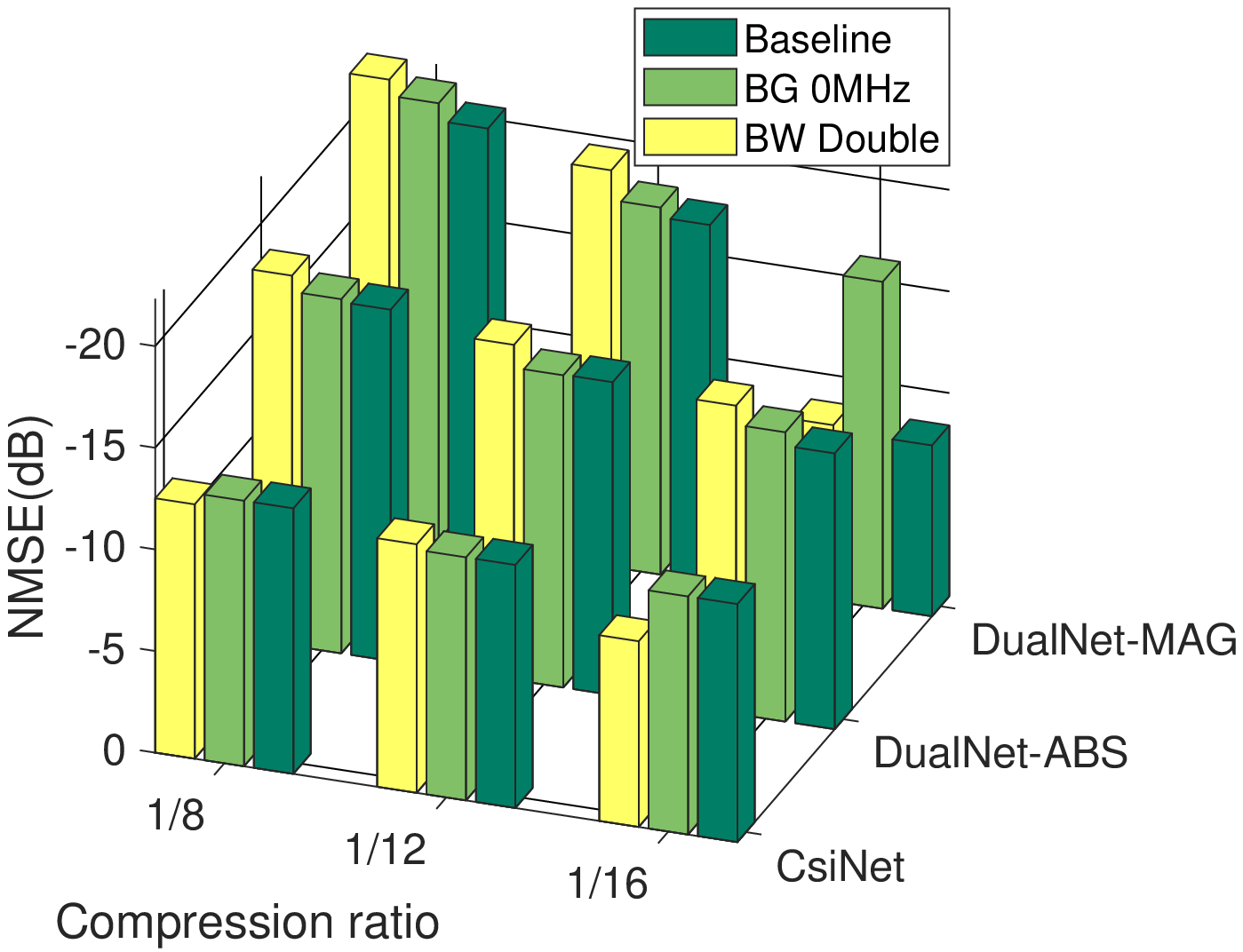}
} 
\subfigure[Outdoor] { \label{fign6:b} 
\includegraphics[width=0.46\columnwidth]{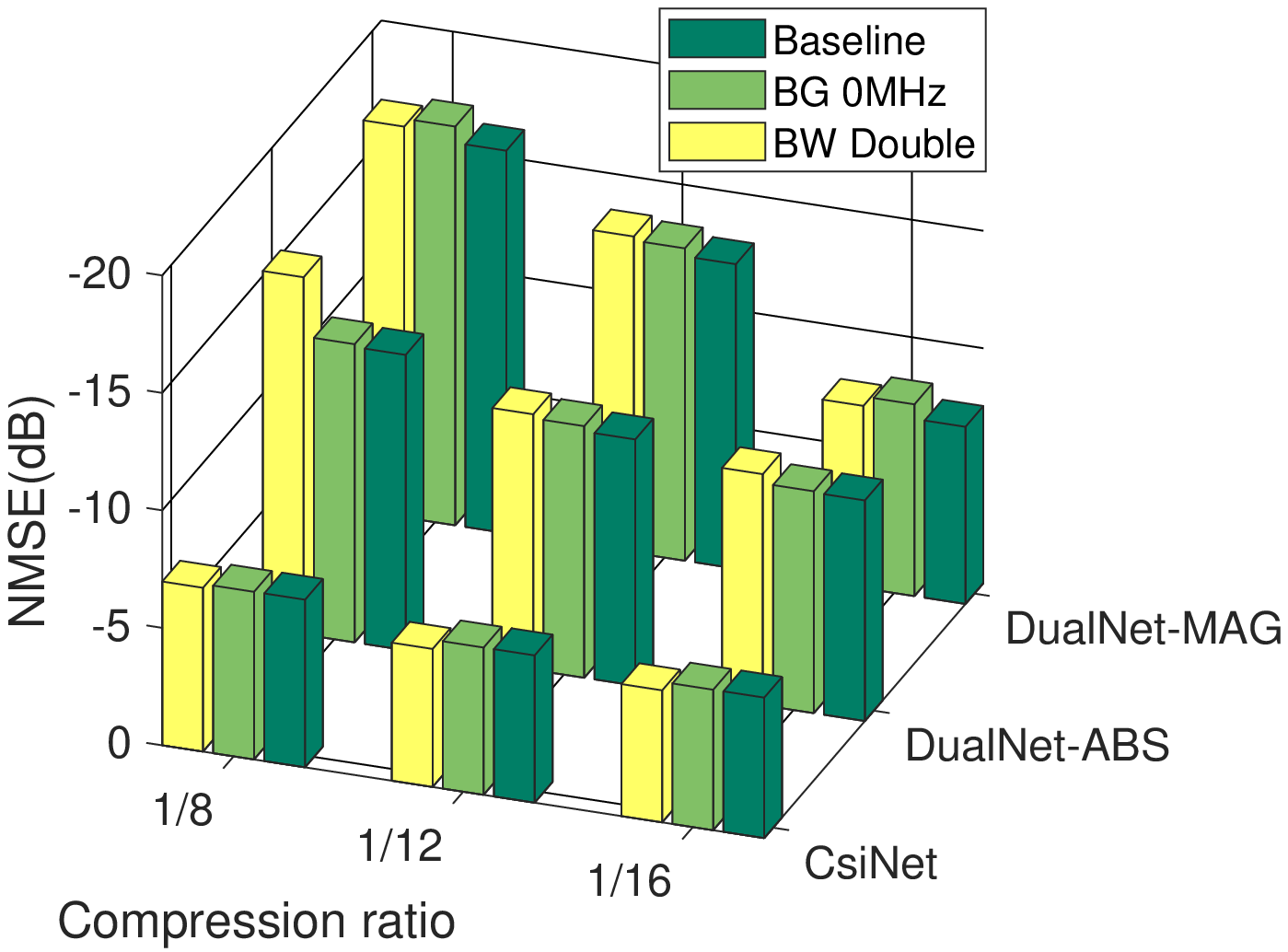} 
} 
\caption{CSI feedback performance in the influence of band gap (BG) and bandwidth (BW).} 
\label{figuren6} \vspace*{-2mm}
\end{figure}

To test the bandwidth effect, we double the channel bandwidth 
by maintaining the same bandgap and test the CSI performance. 
To test the bandgap effect, we reduce the band gap to $0$ 
without changing the bandwidth. From Fig. \ref{figuren6}, 
both DualNet-ABS and DualNet-MAG can improve the downlink CSI recovery accuracy
by reducing the bandgap and/or increasing  the bandwidth.
The performance of CsiNet remains unchanged for different bandgap values,
since  uplink CSI is not used in this method. 
On the other hand, increasing channel bandwidth can
worsen the CsiNet performance in some cases.
One possible reason is that wider bandwidth leads to more uniform CSI distribution, 
thereby requiring larger feedback to achieve 
the same performance.

\subsection{Specialized DL Models Combined With Channel Features}
More specialized DL architectures for wireless communications are still 
under study and should be developed to incorporate wireless channel features
and other domain knowledges.
Most existing DL networks in physical layer of wireless communications
directly adopt known DL architecture or algorithms.
Application specific architectures for wireless networks
would depend on channel models,
signal features, redundancy, and other domain specific knowledge. 


\begin{figure} \centering 
\subfigure[Indoor] {\label{fig5:a} 
\includegraphics[width=0.46\columnwidth]{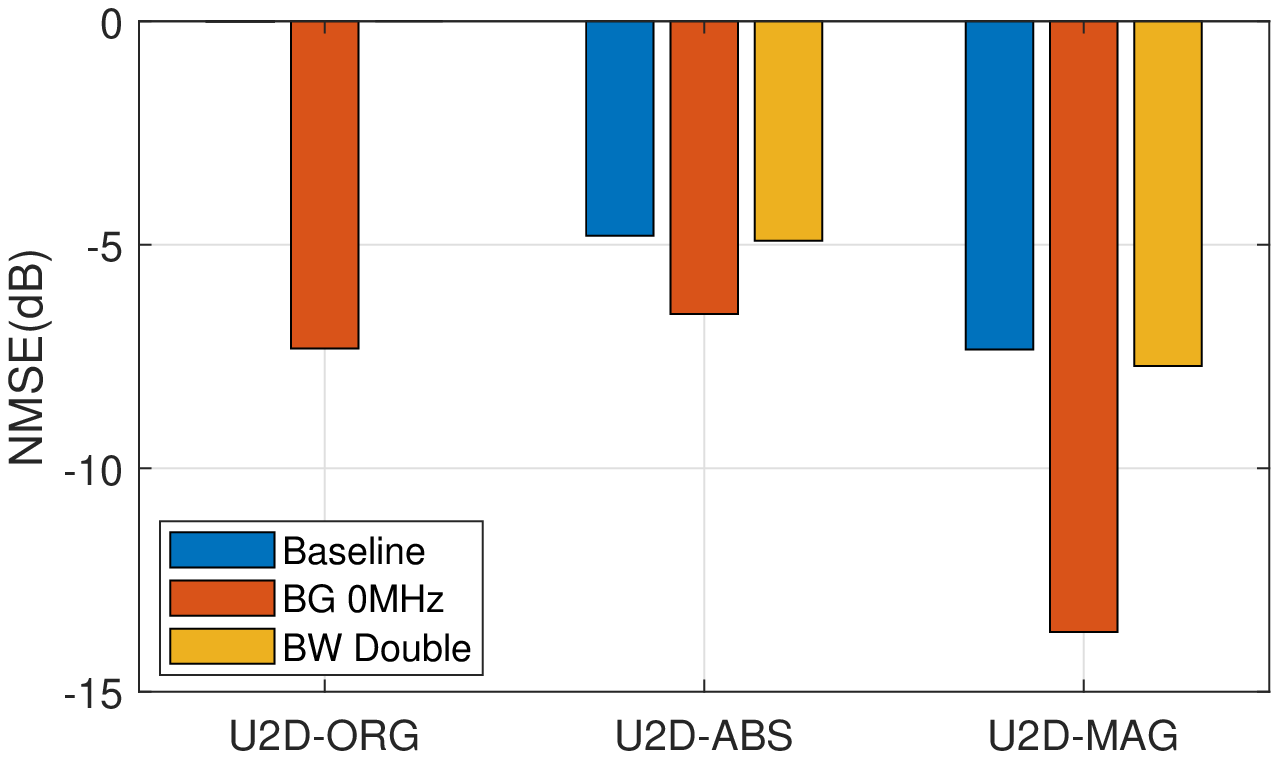}
} 
\subfigure[Outdoor] { \label{fig5:b} 
\includegraphics[width=0.46\columnwidth]{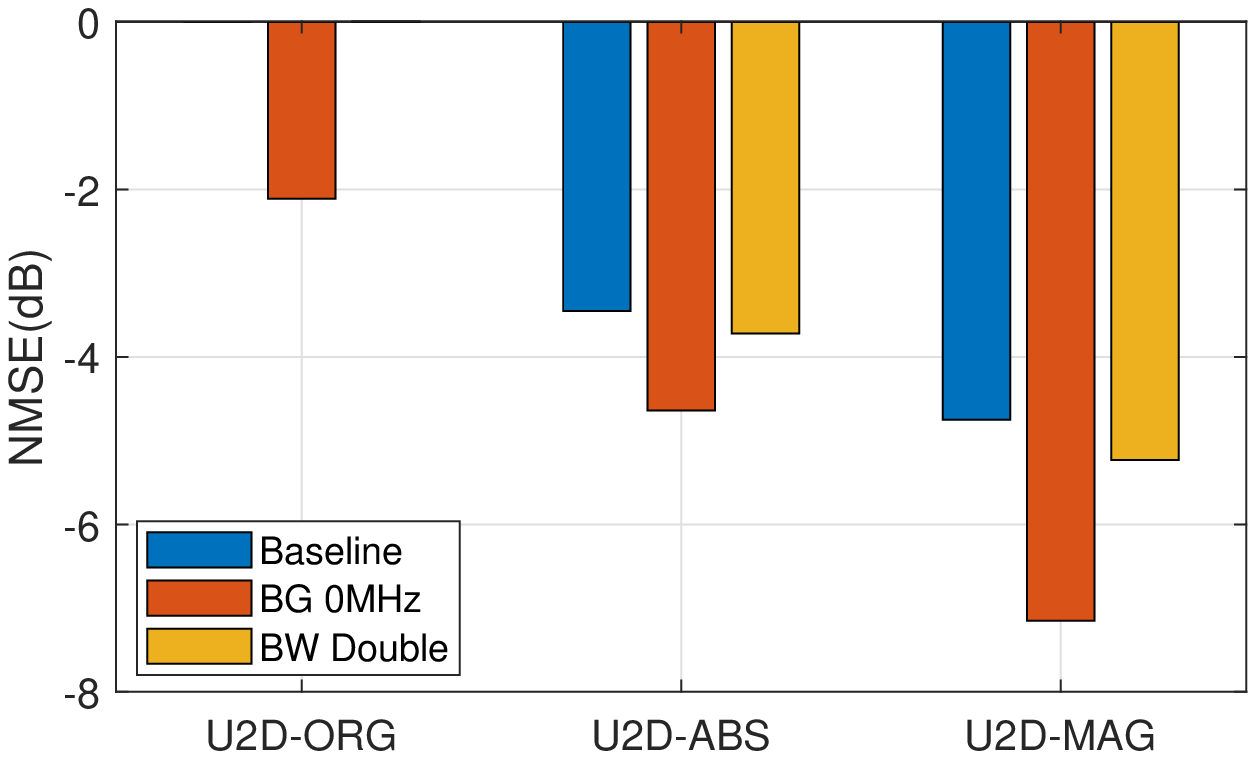} 
} 
\caption{Performance comparison for different bandgap (BG) and bandwidth (BW).} 
\label{figure5} \vspace*{-2mm}
\end{figure}

As an example for combining bi-directional CSI correlation to infer the downlink CSI using uplink CSI,
three architectures are tested \footnote{Details and related codes are provided on: https://ieee-collabratec.ieee.org/app/workspaces/6247/activities}. Unlike DualNet-ABS and DualNet-MAG, U2D-ABS and U2D-MAG respectively recover the absolute values 
and magnitudes of downlink CSI from the corresponding uplink CSI directly without feedback. 
U2D-ORG divides the downlink CSI into real and imaginary parts without separating their signs as the DL network input.
Using the same data set of Section V.A, we consider perfect knowledge of phases and signs of downlink CSI 
at gNB. The performance of Fig. \ref{figure5} shows improved downlink CSI accuracy from U2D-ABS and U2D-MAG
for reduced bandgap and increased bandwidth. U2D-MAG is superior in all cases. By reducing the feedback payload, 
U2D-ABS and U2D-MAG show greater promise for better downlink CSI estimation efficiency.


\subsection{Low Complexity and Distributed DL in Wireless Nodes}
Although DL is a powerful tool in  
CSI estimation, its computation burden
is significant if running on individual
UE node. Considering the battery capacity of mobile devices and  wireless sensors in age of ubiquitous connections, it is vital to reduce 
the complexity of DL algorithms for wireless communications.
An important alternative is to distribute the DL computation load among multiple cooperative nodes.
In wireless communication systems, learning tasks 
can be potentially carried out distributively. 
Distributed DL implementation brings a host of unique challenges such as
task scheduling, node coordination, communications for
data exchange and result transfer, and robustness, 
among others. 

\subsection{Tradeoff between Performance and Training Efficiency}

Existing works and designs have showcased the power of data-driven models in CSI estimation for massive MIMO 
communications. Even though a universal transmitter/receiver can be optimized 
in the end-to-end learning-based design, the training process 
may take very long since results 
from many communication system blocks are merged. 
For practical applications, we need to carefully design 
DL networks and develop algorithms to achieve good tradeoff between
the training efficiency and overall performance. In order to improve 
high training efficiency and achieve good CSI estimation, 
subsets of communication blocks may be pre-calibrated 
and model-driven DL methods can be considered. 


\subsection{Transfer Learning Based Approaches}
Transfer learning allows knowledge 
learned from one task to be transfered to another similar task. 
By avoiding model learning from scratch for each new 
massive MIMO configuration (e.g., antenna number, bandwidth, and mobility), 
transfer learning can shorten the training process in new configuration
and allows DL networks to achieve good CSI
estimates even without access 
or time to use too much training data. 
Practically, there are many active UEs
and gNBs in wireless networks. 
Therefore,
transfer learning is a potential direction for the practical implementation
of DL models in wireless communication networks 
by levering prior knowledges. 

\section{Conclusion}

DL has been
recently emerged as an exciting design tool in
developing future wireless communication systems. 
In this paper, we introduce the
basic principles of applying DL
for improving RF wireless network performance through
the integration of underlying physical channel characteristics 
in practical massive MIMO deployment. 
We provide important insights on how DL benefits from 
physical RF channel properties and present a 
comprehensive overview on the application of DL 
for accurately estimating CSI in massive MIMO communications. 
We provide examples of successful
DL application in CSI estimation and feedback for massive MIMO
wireless systems and outline several
promising directions for future research.


%

\addtolength{\textheight}{-12cm}   

\ifCLASSOPTIONcaptionsoff
  \newpage
\fi

\bibliographystyle{IEEEtran}




\end{document}